\documentclass[a4paper,fleqn,usenatbib]{mnras}

\usepackage{times,txfonts}

\usepackage[T1]{fontenc}
\usepackage{ae,aecompl}

\usepackage{graphicx}	

\title[Molecular gas in galaxies at z $\sim$ 2]{The AT-LESS CO(1-0) survey of submillimetre galaxies in the Extended Chandra Deep Field South: First results on cold molecular gas in galaxies at z $\sim$ 2}

\author[M. T. Huynh et al.]{
Minh T. Huynh,$^{1,2}$\thanks{E-mail: minh.huynh@uwa.edu.au}
B.H.C. Emonts,$^{3}$ A.E. Kimball,$^{4,5}$ N. Seymour,$^6$  Ian Smail,$^7$  \newauthor  
A.M. Swinbank$^7$ W.N. Brandt,$^{8,9,10}$ C.M. Casey,$^{11}$ S.C. Chapman,$^{12}$ H. Dannerbauer,$^{13,14}$  \newauthor 
J.A. Hodge,$^{15}$ R.J. Ivison,$^{16,17}$ E. Schinnerer,$^{18}$ A.P. Thomson,$^{7}$ P. van der Werf,$^{15}$ \newauthor J.L. Wardlow$^{7}$\\
$^{1}$ International Centre for Radio Astronomy Research, M468, University of Western Australia, Crawley, WA 6009, Australia \\
$^{2}$ CSIRO Astronomy and Space Science, 26 Dick Perry Avenue, Kensington WA 6151, Australia \\
$^{3}$ Centro de Astrobiolog{\'i}a (INTA-CSIC), Ctra de Torrej{\'o}n a Ajalvir, km 4, E-28850 Torrej{\'o}n de Ardoz, Madrid, Spain \\
$^{4}$ CSIRO Astronomy and Space Science, PO Box 76, Epping, NSW, 1710, Australia \\
$^{5}$ National Radio Astronomy Observatory, 1003 Lopezville Rd, Socorro, NM, 87801, USA \\
$^{6}$ International Centre for Radio Astronomy Research, Curtin University, Bentley, WA 6102, Australia \\
$^{7}$ Centre for Extragalactic Astronomy, Department of Physics, Durham University, South Road, Durham DH1 3LE UK \\
$^{8}$ Department of Astronomy and Astrophysics, 525 Davey Lab, The Pennsylvania State University, University Park, PA 16802, USA \\
$^{9}$ Institute for Gravitation and the Cosmos, The Pennsylvania State University, University Park, PA 16802, USA \\
$^{10}$ Department of Physics, The Pennsylvania State University, University Park, PA 16802, USA \\
$^{11}$ Department of Astronomy, the University of Texas at Austin, 2515 Speedway Blvd, Stop C1400, Austin, TX 78712, USA \\
$^{12}$ Dalhousie University, Halifax, Nova Scotia B3H 3J5, Canada \\
$^{13}$ Instituto de Astrof{\'i}sica de Canarias (IAC), E-38205 La Laguna, Tenerife, Spain \\
$^{14}$ Universidad de La Laguna, Dpto. Astrof{\'i}sica, E-38206 La Laguna, Tenerife, Spain \\
$^{15}$ Leiden Observatory, Leiden University, PO Box 9513, NL-2300 RA Leiden, the Netherlands \\
$^{16}$ European Southern Observatory, Karl-Schwarzschild-Str. 2, D-85748 Garching bei M{\"u}nchen, Germany \\
$^{17}$ Institute for Astronomy, University of Edinburgh, Royal Observatory, Blackford Hill, Edinburgh EH9 3HJ, UK \\
$^{18}$ Max-Planck-Institut f{\"u}r Astronomie, K{\"o}nigstuhl 17, D-69117 Heidelberg, Germany }

\date{Accepted XXX. Received YYY; in original form ZZZ}

\pubyear{2017}

\begin{document}
\label{firstpage}
\pagerange{\pageref{firstpage}--\pageref{lastpage}}
\maketitle

\begin{abstract}
We present the first results from our on-going Australia Telescope Compact Array survey of $^{12}$CO(1-0) in ALMA-identified submillimetre galaxies in the Extended {\it Chandra} Deep Field South. 
Strong detections of  $^{12}$CO(1-0) emission from two submillimetre galaxies, ALESS 122.1 ($z = 2.0232$) and ALESS 67.1 ($z = 2.1230$), were obtained. We estimate gas masses of $M_{\rm gas} \sim 1.3 \times 10^{11}$ ${\rm M}_\odot$ and $M_{\rm gas} \sim 1.0 \times 10^{11}$${\rm M}_\odot$ for ALESS 122.1  and ALESS 67.1, respectively, adopting $\alpha_{\rm CO} = 1.0$. Dynamical mass estimates from the kinematics of the $^{12}$CO(1-0)  line yields $M_{\rm dyn} \sin^2 i$ = (2.1 $\pm$ 1.1) $\times 10^{11}$ M$_\odot$ and (3.2 $\pm$ 0.9) $\times 10^{11}$ M$_\odot$ for ALESS 122.1 and ALESS 67.1, respectively. This is consistent with the total baryonic mass estimates of these two systems. 
We examine star formation efficiency using the $L_{\rm FIR}$ versus $L'_{\rm CO(1-0)}$ relation for samples of local ULIRGs and LIRGs, and more distant star-forming galaxies, with $^{12}$CO(1-0) detections. We find some evidence of a shallower slope for ULIRGs and SMGs compared to less luminous systems, but a larger sample is required for definite conclusions. We determine gas-to-dust ratios of 170 $\pm$ 30 and 140 $\pm$ 30 for ALESS 122.1 and ALESS 67.1, respectively, showing ALESS 122.1 has an unusually large gas reservoir. By combining the 38.1 GHz continuum detection of ALESS 122.1 with 1.4 and 5.5 GHz data, we estimate that the free-free contribution to radio emission at 38.1 GHz is 34 $\pm$ 17 $\mu$Jy, yielding a star formation rate (1400 $\pm$ 700 ${\rm M}_\odot$ yr$^{-1}$) consistent with that from the infrared luminosity. 

\end{abstract}

\begin{keywords}
galaxies: evolution -- submillimetre: galaxies -- radio lines: galaxies
\end{keywords}



\section{Introduction}

Since their initial discovery, submillimeter galaxies (SMGs) have become an important element of our understanding of cosmic galaxy formation and evolution (e.g. \citealp{blain2002, casey2014}). 
Selected in the rest-frame far-infrared (FIR), SMGs contain significant masses of cold dust \citep{casey2012} and large reservoirs of molecular gas ($\gtrsim 10^{10}$ M$_\odot$; e.g. \citealp{bothwell2013}).  These luminous galaxies have median redshifts of $z \sim 2$~--~3 \citep{chapman2005,wardlow2011,smolcic2012,simpson2014}, and extreme far-infrared (FIR) luminosities $L_{FIR} > 10^{12} L_\odot$  implying large star formation rates of  $\sim$ 100~--~1000 $\rm{M}_\odot$ yr$^{-1}$ (e.g. \citealp{blain2002}). 
The peak of star formation in the Universe also occurred at $z \sim 2$~--~3 \citep{ahopkins2006,madau2014} and ultraluminous infrared galaxies (ULIRGs) are responsible for roughly half of the cosmic infrared luminosity density at those redshifts \citep{magnelli2013,gruppioni2013}, suggesting SMGs play a vital role in galaxy evolution. 
 
Since the extreme star formation rates of local ULIRGs are believed to be driven by major mergers, it has also been asserted that SMGs at high redshift have similar evolutionary histories (e.g. \citealp{engel2010,chencc2015}). However, secular origins have also been proposed for SMGs, one justification being that there are not enough mergers in some simulations to account for the number of observed SMGs (e.g. \citealp{hopkins2010,hayward2013}). Cosmological hydrodynamic simulations are now able to reproduce some SMG properties, such as stellar masses and molecular gas fractions, without a major merger (e.g \citealp{dave2010, narayanan2015}), but they in general can not reproduce the highest SFRs seen in SMGs or match the properties of descendants at $z \sim 0$. Moreover, the SMG population appears to be diverse and so large observational samples are necessary to capture this diversity and test the models. 

Molecular gas studies of SMGs provide unique insight into the physical properties of these systems. 
Molecular line observations provide information on the kinematics of the galaxy (i.e. turbulent versus ordered rotating discs), as well as dynamical and gas mass estimates. 
Radio emission produced by the rotational transition of carbon monoxide ($^{12}$CO) is one of the most accessible tracers of cold molecular gas in galaxies \citep{carilli2013}.
However only a few tens of unlensed SMGs at $z \gtrsim 1.5$ have been detected in CO (e.g. {\citealp{carilli2013}) and most are of high J (J $ > 2$) transitions. 
The high-CO transitions trace dense and thermally excited gas in the starburst/AGN regions, while only the lowest CO transitions fully reveal the more widely distributed reservoirs of less dense, sub-thermally excited gas (e.g. \citealp{papadopoulos2000, papadopoulos2001, carilli2010,ivison2011}). The ground transition CO(1-0) is least affected by the excitation conditions of the gas and therefore provides the most robust estimates of overall molecular gas content and the broadest tracer of the dynamics of the system. 

To study the molecular gas content of SMGs we have initiated a survey of CO(1-0) with the the Australia Telescope Compact Array (ATCA). We describe the SMG sample and the ATCA observations in Section 2. 
The results of the observations are presented in Section 3. In Section 4 we discuss the molecular gas masses, dynamical masses, star formation efficiency and dust-to-gas ratios of the observed systems. 
The standard $\Lambda$-CDM cosmological parameters of $\Omega_{\rm M} = 0.29$,  $\Omega_{\rm \Lambda} = 0.71$, and a Hubble constant of 70 km s$^{-1}$ Mpc$^{-1}$ are adopted throughout this paper. 

\section{Observations and Data Reduction}

\subsection{Our ALESS SMG Sample}

 Our sample is selected from the ALMA study of 99 submillimeter sources from the ALMA LABOCA Extended {\it Chandra} Deep Field South (ECDFS) Submillimeter Survey (ALESS, \citealp{weiss2009,hodge2013}).  
ALESS is an ALMA Cycle 0 survey at 870 $\mu$m to follow up 122 of the original 126 submm sources detected by the LABOCA ECDFS Submillimeter Survey (LESS, \citealp{weiss2009}). The excellent angular resolution and sensitivity of ALMA ($\sim$1.5 arcsec and about 3 times deeper than LESS) resulted in a sample of 99 statistically reliable SMGs \citep{hodge2013}. This large ALMA-identified SMG sample is free of the biases and misidentifications which have affected previous SMG studies.  
 
 The ECDFS (RA = 03h32m28s, Dec = $-$27$^\circ$48$'$30$''$) is one of the best-studied extragalactic survey fields available, allowing for secure identification of counterparts at other wavelengths. A spectroscopic survey of the original LESS SMGs was performed as part of a VLT Large Programme with the FOcal Reducer
and low dispersion Spectrograph (FORS2) and VIsible MultiObject Spectrograph (VIMOS) during 2009 -- 2012 (Danielson et al. 2016, submitted). To supplement the Large Programme, and target ALMA-identified ALESS SMGs which differed from the original LESS counterparts,  observations were also obtained on XSHOOTER on the VLT, Gemini Near-Infrared Spectrograph (GNIRS) on Gemini South, the Multi-Object Spectrometer for Infra-Red Exploration (MOSFIRE) on Keck I, and DEep Imaging Multi-Object Spectrograph
(DEIMOS) on Keck II. This extensive spectroscopic campaign has provided secure redshifts for 51/99 ALESS SMGs (Danielson et al. 2016, submitted).  {\em Herschel} SPIRE imaging was deblended by combining the ALMA, {\em Spitzer} Multiband Imaging Photometer (MIPS) 24 $\mu$m and radio catalogue priors, to determine far-infrared properties, including dust masses, total infrared luminosities and star formation rates of the individual SMGs \citep{swinbank2014}. 

Our initial sample consists of nine ALESS SMGs at $1.5 < z < 2.5$, where CO(1-0) is detectable in the Australia Telescope Compact Array (ATCA) 7mm band, with the best quality optical spectra from Danielson et al. (2016). These nine SMGs have secure redshifts with multiple emission or absorption features identified in the optical spectra. Pilot ATCA observations were granted in the OCT2015 semester in which we targeted two ALESS SMGs: ALESS 122.1 and ALESS 67.1. These two targets were chosen as they are the most infrared luminous SMGs out of the nine, and hence most likely to have detectable molecular gas reservoirs. 

\subsection{ATCA Observations and Data Reduction} 

Observations of ALESS122.1 and ALESS67.1 were performed on the Australia Telescope Compact Array (ATCA), using the Compact Array Broadband Backend (CABB; \citealp{wilson2011}), in August and September 2015. The array was in the standard compact hybrid configurations H75 and H168 for ALESS 122.1 and ALESS 67.1, resulting in maximum baselines of 89m and 192m, respectively, discarding the outer 6th antenna. The hybrid configurations, consisting of two antennas along the northern spur, allow good $(u,v)$ coverage to be obtained for integrations less than the full 12 hour synthesis. Our observations consisted of $\sim$8 hour runs to ensure the source elevation is greater than $\sim$30 degrees. We obtained total integration times of 38 and 31 hours on-source for  ALESS122.1 and ALESS67.1, respectively.  The weather was good to average, with atmospheric path length rms variations generally in the range of 50 to 400 $\mu$m, as measured on the 230m baseline ATCA Seeing Monitor \citep{middelberg2006}. The 7mm receiver was centred at the expected frequency of the $^{12}$CO(1-0) line emission ($\nu_{\rm rest}$ = 115.2712 GHz) given their spectroscopic redshifts, i.e. 38.129 GHz for ALESS122.1 and 36.910 GHz for ALESS67.1 GHz. The 2GHz bandwidth of CABB results in a velocity coverage of approximately 15,000 km s$^{-1}$.

Following \cite{emonts2011}, a bandpass calibration scan was acquired at the beginning and end of each 8 hour night; however we found that the bandpass scan in the beginning of the night is sufficient for good bandpass calibration and the second scan is thus a backup.  Phase and amplitude calibration information was acquired with 2 minute scans on PKS 0346$-$279 every 15 minutes and pointing checks performed on the same source every hour. For flux calibration we observed Uranus at a time when it was close to the same elevation as our targets; this occurred around 00:30 LST and at an elevation of $\sim$50 degrees. The uncertainty in the flux density calibration using the standard {\sc miriad} model of Uranus is estimated to be 30\% \citep{emonts2011}, but can be as little as 20\% when following this scheme and in good conditions. 

The data were calibrated, mapped and analysed using the standard {\sc miriad} \citep{sault1999} and {\sc karma} \citep{gooch1996} packages. 
The synthesized beam from natural weighting is 14.4 $\times$ 10.6 arcsec and 7.0 $\times$ 4.6 arcsec for ALESS 122.1 and ALESS 67.1, respectively. 
The resultant noise in the single 1 MHz ($\sim$8 km s$^{-1}$) channels is $\sim$ 0.40 mJy beam$^{-1}$ for both the  ALESS122.1 and ALESS 67.1 cubes, consistent with other comparable 7mm ATCA/CABB studies (e.g. \citealp{coppin2010, emonts2014, huynh2014, emonts2015}).

\section{Results}

The visibilities were re-sampled to produce cubes with velocity resolutions of 200, 300, 400 and 600 km s$^{-1}$ and each cube was examined for an emission line at the expected redshift and centered near the ALMA position. 
We identify a line at the ALMA position and spectroscopic redshift in the cubes for both sources at more than 8$\sigma$ significance, and across multiple channels for 200 and 300 km~s$^{-1}$ binning. 
The 200 km~s$^{-1}$ binned spectra have similar sensitivities (0.079 mJy/beam) and the CO peaks are detected at $\sim$12$\sigma$ and $\sim$8$\sigma$ significance in the brightest channel for ALESS 122.1 and ALESS 67.1, respectively (see Figure \ref{fig:spectra}). 
The CO emission is coincident with the ALMA submillimeter source and has a clear IRAC counterpart (Figure \ref{fig:stamps}).

Gaussian fits were performed on the 200 km s$^{-1}$ cube to obtain $^{12}$CO(1-0) line parameters, which are summarised in Table 1. The CO(1-0) spectrum for ALESS122.1 has a significant fitted continuum of 70 $\pm$ 20 $\mu$Jy, which is subtracted from the spectra in Figure 1. The CO(1-0) line for ALESS 122.1 then has a fitted peak of 0.86 $\pm$ 0.06 mJy, a FWHM of 700 $\pm$ 60 km~s$^{-1}$.  The CO(1-0) line for ALESS 67.1 has a fitted peak of 0.58 $\pm$ 0.07 mJy and a FWHM of 710 $\pm$ 90 km~s$^{-1}$. Both lines are consistent with having a zero velocity offset with respect to the optical spectroscopic redshift. Integrating the best fit Gaussian yields a line luminosity of 0.64 $\pm$ 0.07 and 0.44 $\pm$ 0.08 Jy km~s$^{-1}$ for ALESS 122.1 and ALESS67.1, respectively. These values do not include the flux calibration uncertainties, which are about 20--30\% \citep{emonts2011}. 

Continuum images were also made from the full CABB 2 GHz bandwidth for each SMG. The central 200 channels were flagged to remove any CO(1-0) flux, and natural weighting used to achieve the highest sensitivity. Both SMGs are detected in the continuum images as point sources, with ALESS 122.1 having a 38.1 GHz flux density of $60 \pm 10$ $\mu$Jy and ALESS 67.1 having a 36.9 GHz flux density of $48 \pm 11$ $\mu$Jy.

\begin{figure*}
\includegraphics[width=8.1cm]{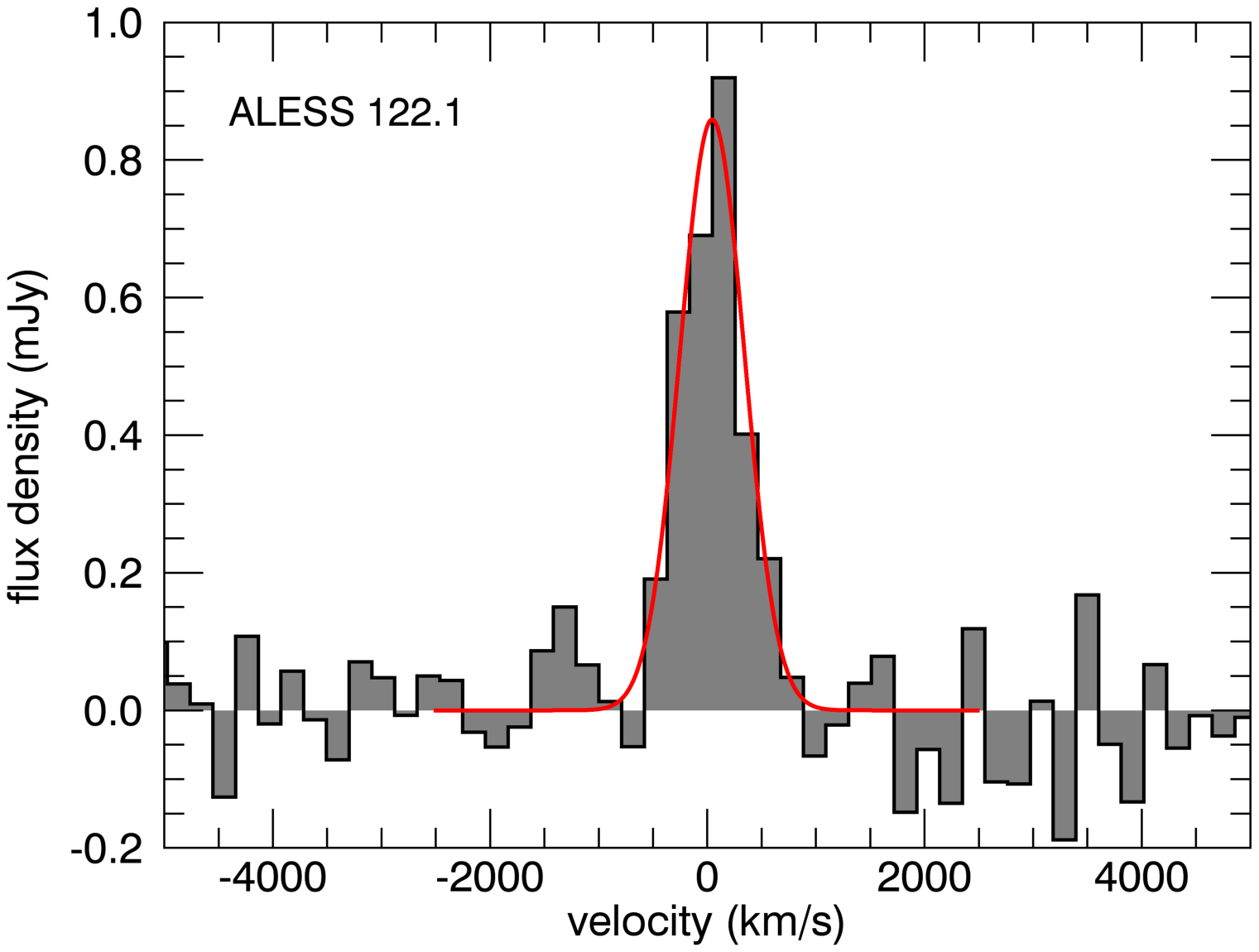}
\hspace{5mm}
\includegraphics[width=8.1cm]{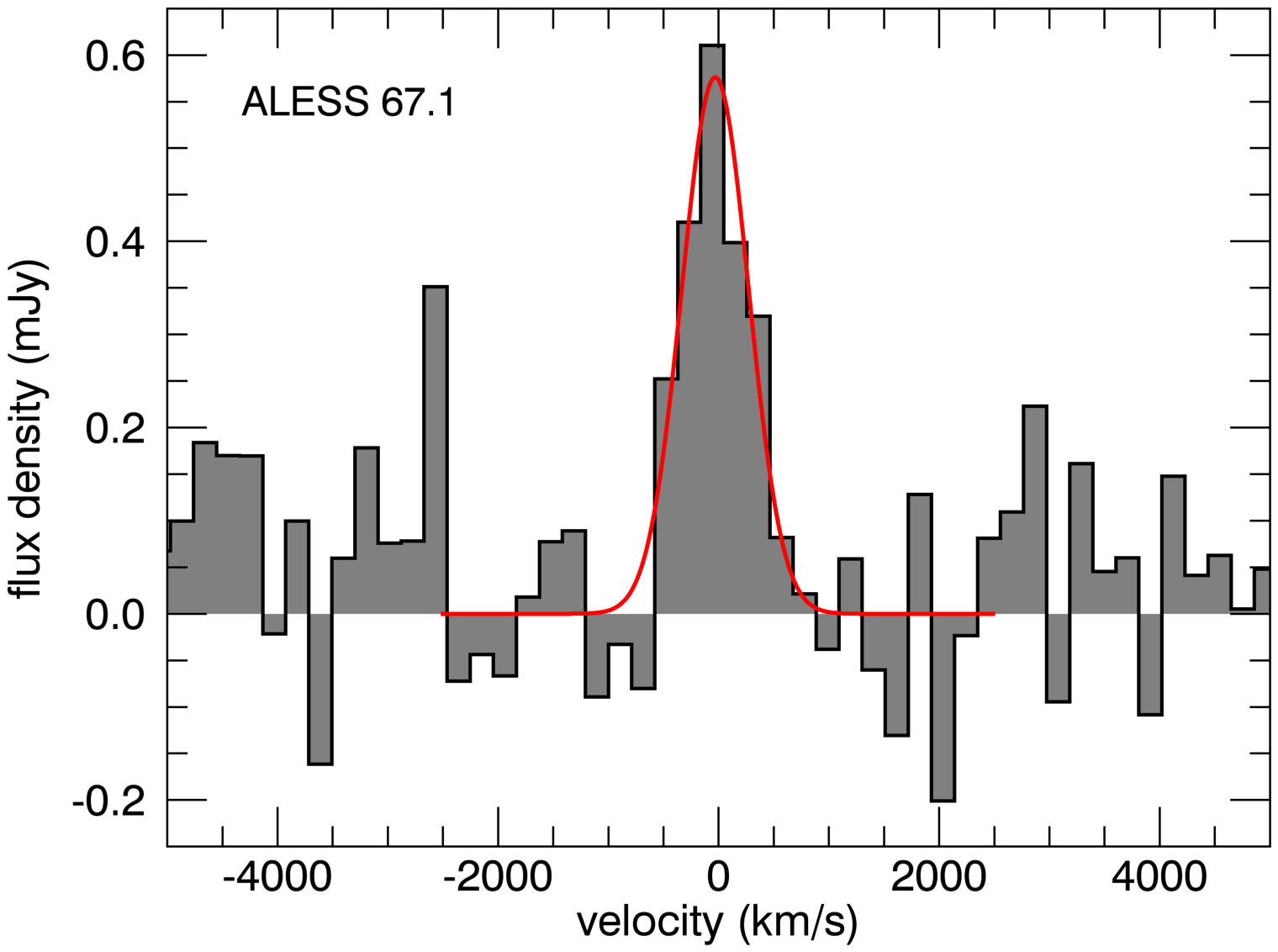}
\caption{The CO(1-0) spectrum for ALESS 122.1 (left, continuum subtracted) and ALESS 67.1 (right), binned to 200 km s$^{-1}$ resolution. The best fit Gaussian is shown as a red line. The CO emission peaks at $\gtrsim$8$\sigma$ significance with this binning with both lines well-fit by single Gaussians at this resolution.}
\label{fig:spectra}
\end{figure*}

\begin{figure*}
\includegraphics[width=6.cm]{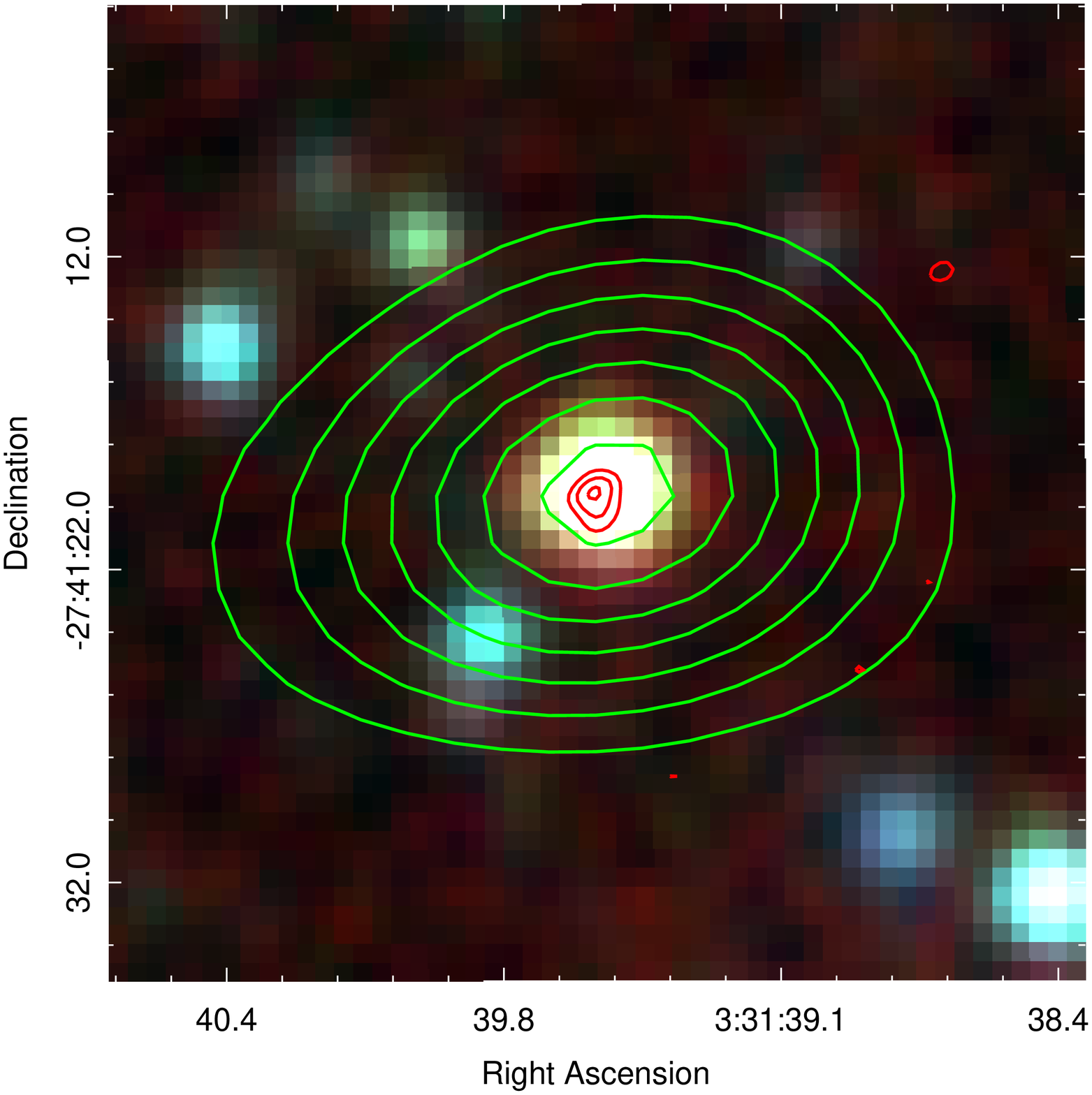}
\hspace{15mm}
\includegraphics[width=6.cm,trim=0 0 1 0,clip]{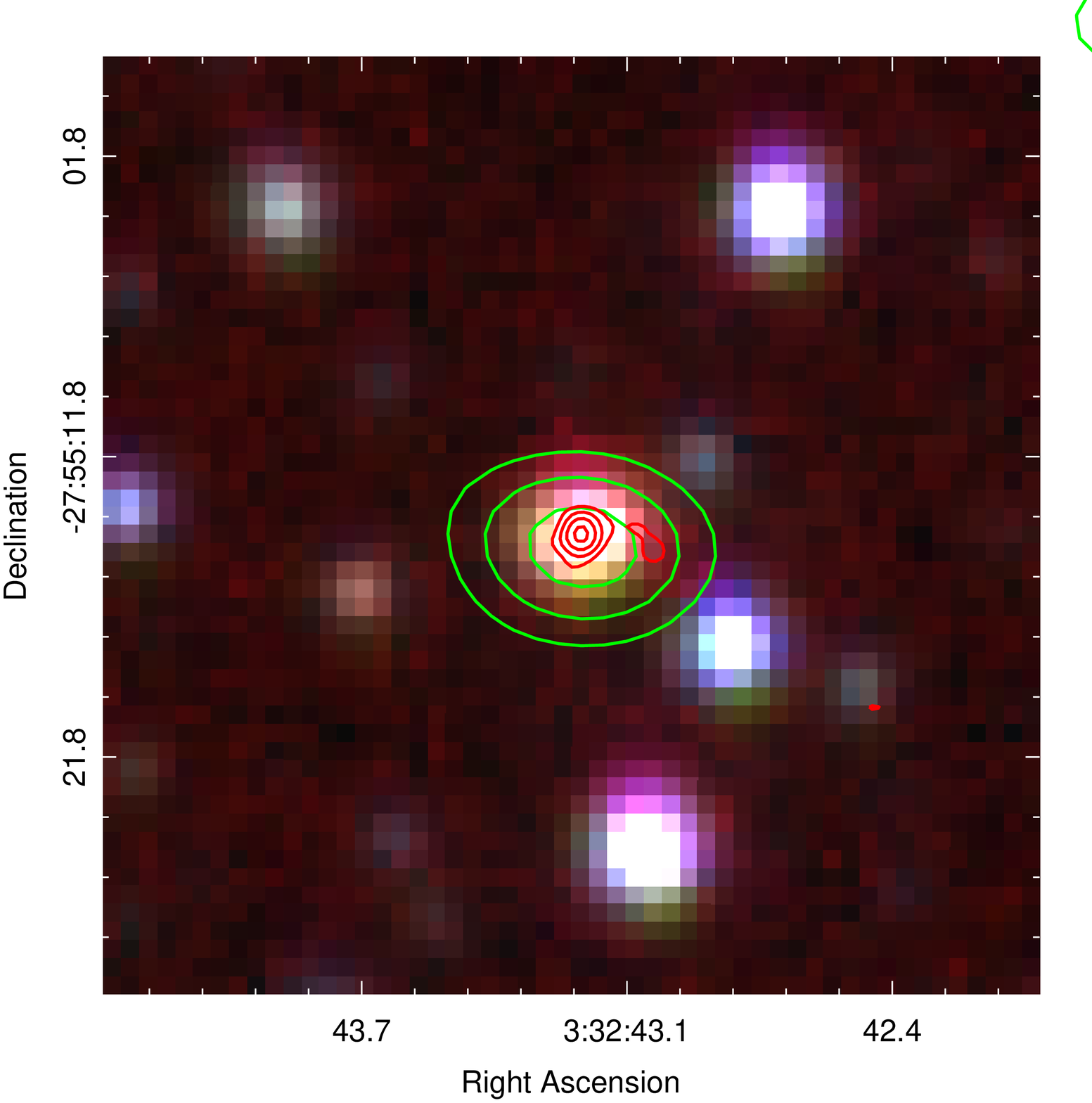}
\caption{False-colour images made from {\em Spitzer} Infrared Array Camera (IRAC) images of \protect\cite{damen2011}\protect\footnotemark[1] for ALESS 122.1(left) and ALESS 67.1 (right). Green contours indicate the CO(1-0) emission from the 600 km s$^{-1}$ cube at 3, 5, 7 .. $\times \; \sigma$. The red contours are from the ALMA 870 $\mu$m continuum map \citep{hodge2013} at  3, 5, 7 .. $\times \; \sigma$. The postage stamp images are 30 $\times$ 30 arcsec.}
\label{fig:stamps}
\end{figure*}

\begin{table}
\renewcommand{\thefootnote}{\alph{footnote}}
\centering
\begin{minipage}{\columnwidth}
\caption{Observed and derived properties of ALESS122.1 and ALESS67.1}
\begin{tabular}{lcc}  \hline
                         & ALESS122.1 & ALESS67.1 \\ \hline
RA$_{\rm CO(1-0)}$ (J2000) & 03:31:39.53 & 03:32:43.20 \\
Dec$_{\rm CO(1-0)}$ (J2000) &  -27:41:19.6 & -27:55:14.8 \\       
$z_{\rm spec}$ & 2.0232  & 2.1230\\
$z_{\rm CO(1-0)}$ & 2.0238 $\pm$ 0.0003 & 2.1228 $\pm$ 0.0004 \\
$L_{\rm FIR}$\footnotemark[1] ($10^{12} L_{\odot}$)& 6.3$^{+0.4}_{-0.5}$ & 5.3$^{+0.7}_{-1.3}$  \\
SFR\footnotemark[2] (M$_\odot$ yr$^{-1}$) & 940$^{+60}_{-80}$ & 790$^{+100}_{-190}$  \\
$M_{\rm dust}$\footnotemark[1] ($10^{8} M_{\odot}$)& 7.9 $\pm$ 0.9 & 7.1 $\pm$ 0.6 \\
Dust temperature\footnotemark[1] (K) & 32  & 31  \\
peak$_{\rm CO(1-0)}$ (mJy) & 0.86 $\pm$ 0.06 & 0.58 $\pm$ 0.07 \\
FWHM$_{\rm CO(1-0)}$ (km s$^{-1}$) & 700 $\pm$ 60  & 710 $\pm$ 90 \\
CO(1-0) line center (km s$^{-1}$) & 45 $\pm$ 30 & $-$31 $\pm$ 40 \\
$I_{\rm CO(1-0)}$ (Jy km s$^{-1}$) & $0.64 \pm 0.07$ &  $0.44 \pm 0.08$ \\  
$L'_{\rm CO(1-0)}$ ($10^{10}$ K km s$^{-1}$ pc$^2$) &  13 $\pm$ 2  &  9.9 $\pm$ 1.8\\
$M$(H$_2$) ($10^{10}$ ${\rm M}_\odot$)\footnotemark[3]  & 13 $\pm$ 2  &  9.9 $\pm$ 1.8\\
$S_{\rm 7mm\; continuum}$ ($\mu$Jy) &  $60 \pm 10$ & $48 \pm 11$ \\
\hline
\end{tabular}
\vspace{-7mm}
\footnotetext[1]{from Swinbank et al. (2014)} 
\footnotetext[2]{using $L_{\rm FIR}$ conversion from Kennicutt and Evans (2012)} 
\footnotetext[3]{Adopting $\alpha_{\rm CO}$ = 1} 
\end{minipage}
\end{table}

\section{Analysis and Discussion}

\subsection{Molecular Gas Masses}

CO detections provide a tool for deriving the masses of the molecular gas reservoirs of these systems. This is important as the reservoir of molecular gas is the raw material from which new stars will be formed, thus giving an indication of the final mass of these systems post the starburst phase (modulo gas falling into or being ejected from the system). 
The ground transition of CO(1-0) is particularly powerful as no assumption of a brightness ratio, or gas spectral line energy distribution (SLED), is required to convert down from a higher $J$ transition. 

Following the method of \cite{solomon2005}, we find ALESS 122.1 has a line luminosity of $L'_{\rm CO(1-0)}$ = (1.3 $\pm$ 0.2) $\times$ $10^{11}$ K km s$^{-1}$ pc$^2$ and ALESS67.1 has a line luminosity of $L'_{\rm CO(1-0)}$ = (9.9 $\pm$ 1.8) $\times$ $10^{10}$ K km s$^{-1}$ pc$^2$ (Table 1). 
A CO-to-H$_2$ conversion factor, $\alpha_{\rm CO}$, is then required to convert the line luminosity to a total molecular gas mass, $M$(H$_2$).
At $z \sim 0$ disc galaxies such as the Milky Way have relatively large values of $\alpha_{\rm CO}$ $\sim$ 3 -- 5 , while a smaller value of $\alpha_{\rm CO}$ = 0.8 is believed to be appropriate for local ULIRGs (e.g. Downes \& Solomon 1998), and this has been widely used for high redshift SMGs. However there is some evidence that the local ULIRG value of $\alpha_{\rm CO} =0.8$ leads to under-estimated gas masses \citep{bothwell2010} and at least one SMG is known to have a large $\alpha$ $\sim$ 2 \citep{swinbank2011, danielson2011}. Given the uncertainty in the conversion factor we adopt $\alpha_{\rm CO} = 1$, following \cite{bothwell2013}, and derive  molecular gas masses, $M$(H$_2$), of (1.3 $\pm$ 0.2) $\times$ $10^{11}$  ${\rm M}_\odot$ and (9.9 $\pm$ 1.8) $\times$ $10^{10}$ ${\rm M}_\odot$ for ALESS 122.1 and ALESS 67.1, respectively.  The mean gas mass for a representative sample of $z \sim 2$ SMGs from $J \gtrsim 3$ observations has been found to be (3.2 $\pm$ 2.1) $\times$ $10^{10}$ ${\rm M}_\odot$ \citep{bothwell2013}. The high-J transitions used in the earlier work may explain some of the discrepancy, but nevertheless ALESS 122.1 and ALESS 67.1 appear to have relatively large gas masses compared to the general SMG population. This is not unexpected as they were selected to have large IR luminosities. 

\footnotetext[1]{https://irsa.ipac.caltech.edu/data/SPITZER/SIMPLE/}

\subsection{CO Line Kinematics and Dynamical Masses}

The kinematics of the CO line traces the gravitational potential well in the system. Previous studies have found very broad CO emission lines from SMGs, but are typically from $J \gtrsim 3$ transition lines.  
\cite{bothwell2013} find a mean FWHM of 510 $\pm$ 80 km s$^{-1}$ for their sample of SMGs.  A greater mean FWHM of 780 km s$^{-1}$  was found for a more infrared-luminous subset of those SMGs \citep{greve2005}. 
More infrared-luminous sources could potentially have a greater dynamical mass, and therefore a larger CO line FWHM. The CO line FWHMs of ALESS 122.1 and ALESS 67.1 are $\sim$700 km s$^{-1}$, which are larger than most of the \cite{bothwell2013} SMG sample, but in line with what is expected from the more infrared-luminous SMGs. The measured CO(1-0) linewidth may also be relatively large due to the $J$ = 1-0 transition being more spatially extended than higher $J$ transitions (e.g. \citealp{ivison2011,emonts2015}).

The width of the CO line from the SMGs allows an estimation of the galaxy dynamical masses, given a spatial extent of the system and an assumption about the dynamical structure. 
Following \cite{solomon2005}, the dynamical mass of the system can be estimated using $M_{\rm dyn} \sin^2 i = 233.5 V^2 R$, where $R$ is the radius of the molecular disk or half the separation between components in a merger model, measured in parsec, and $V$ is the FWHM of the CO line or half the separation in velocity of the component CO lines in a merger model, measured in km s$^{-1}$. 
The HST images \citep{rix2004} show various clumps that indicate ALESS 122.1 and ALESS 67.1 could be mergers (Figure \ref{fig:redblue}). Taking the merger model and $R$ to be $\lesssim$0.5 arcsec (4.2 kpc at $z = 2$), which is the approximate separation of the clumps in HST optical counterparts  (Figure \ref{fig:redblue}), the dynamical mass, $M_{\rm dyn} \sin^2 i$, is $\lesssim$$5 \times 10^{11}$ M$_\odot$ for both ALESS 122.1 and ALESS67.1.

However, the HST image does not reveal whether or not the systems contain a gaseous disc. In the presence of a gas-disc, the dynamical mass can be estimated from the CO line kinematics, by comparing the spatial offset between the redshifted and blueshifted components of the CO line. This was determined by making 300 km s$^{-1}$ wide channel maps centred at $-300$ to 0 km s$^{-1}$ and $0$ to +300 km s$^{-1}$ (Figure \ref{fig:redblue}). The centroid of the CO emission was determined for the maps, and we derive a spatial offset between the redshifted and blueshifted emission of 1.2 $\pm$ 0.6 arcsec (10.2 $\pm$ 5.1 kpc)  and 1.8$\pm$ 0.5 arcsec (15.2 $\pm$ 4.2 kpc) for ALESS122.1 and ALESS67.1, respectively. 
This is comparable to CO(1-0) sizes observed in other SMGs by \cite{ivison2011}.
Assuming this spatial offset represents rotating gas, then the dynamical mass $M_{\rm dyn} \sin^2 i$ = (2.1 $\pm$ 1.1) $\times 10^{11}$ M$_\odot$ and (3.2 $\pm$ 0.9) $\times 10^{11}$ M$_\odot$ for ALESS 122.1 and ALESS 67.1, respectively. These estimates are consistent with the dynamical masses derived from using the CO line FWHM and optical size from HST imaging. For a rotating model, \cite{bothwell2013} found their sample of SMGs to have a median dynamical mass of (1.6 $\pm$ 0.3)  $\times$ 10$^{10}$ $R$ M$_\odot$, where radius $R$ is in kpc. Using the measured spatial offsets of 10 and 15 kpc for $R$, the dynamical masses of ALESS 122.1 and ALESS 67.1 are consistent with the median dynamical mass of their SMG sample. 

The total baryonic mass can be calculated by combining the gas and stellar mass estimates for the SMGs. The stellar masses of ALESS 122.1 and ALESS 67.1 have been estimated by \cite{simpson2014}, who used absolute restframe $H$-band magnitudes from best fit SEDs to multi-band photometry and a model mass-to-light ratio. They estimate ALESS 122.1 has a stellar mass of $5.2 \times 10^{10}$ M$_\odot$  and ALESS 67.1 has a stellar mass of $4.4 \times 10^{10}$ M$_\odot$. These estimates are uncertain by a factor of at least a few, due to the difficulty in distinguishing between different model star formation histories to predict an accurate mass-to-light ratio \citep{hainline2011}. Nevertheless, this suggests the total baryonic mass of the system, $M_{\rm bary} = M_{\rm H_2} + M_{\rm stars}$, is then (1.9 $\pm$ 1.1) $\times$ 10$^{11}$ M$_\odot$ and  (1.4 $\pm$ 0.9) $\times$ 10$^{11}$ M$_\odot$ for ALESS 122.1 and ALESS 67.1, respectively, where the uncertainties for the stellar mass estimate is assumed to be a factor of two. These mass estimates are consistent with the derived dynamical masses, given the uncertainties involved.  Given these total baryonic mass estimates, ALESS 122.1 and ALESS 67.1 have gas mass fractions of $\sim$70\%, which, although large, have been been observed in other gas-dominated SMGs (e.g. \citealp{bothwell2013}) and $z \sim 2$ UV-selected massive star forming galaxies \citep{tacconi2010}.

\begin{figure*}
\includegraphics[width=6.cm,angle=270]{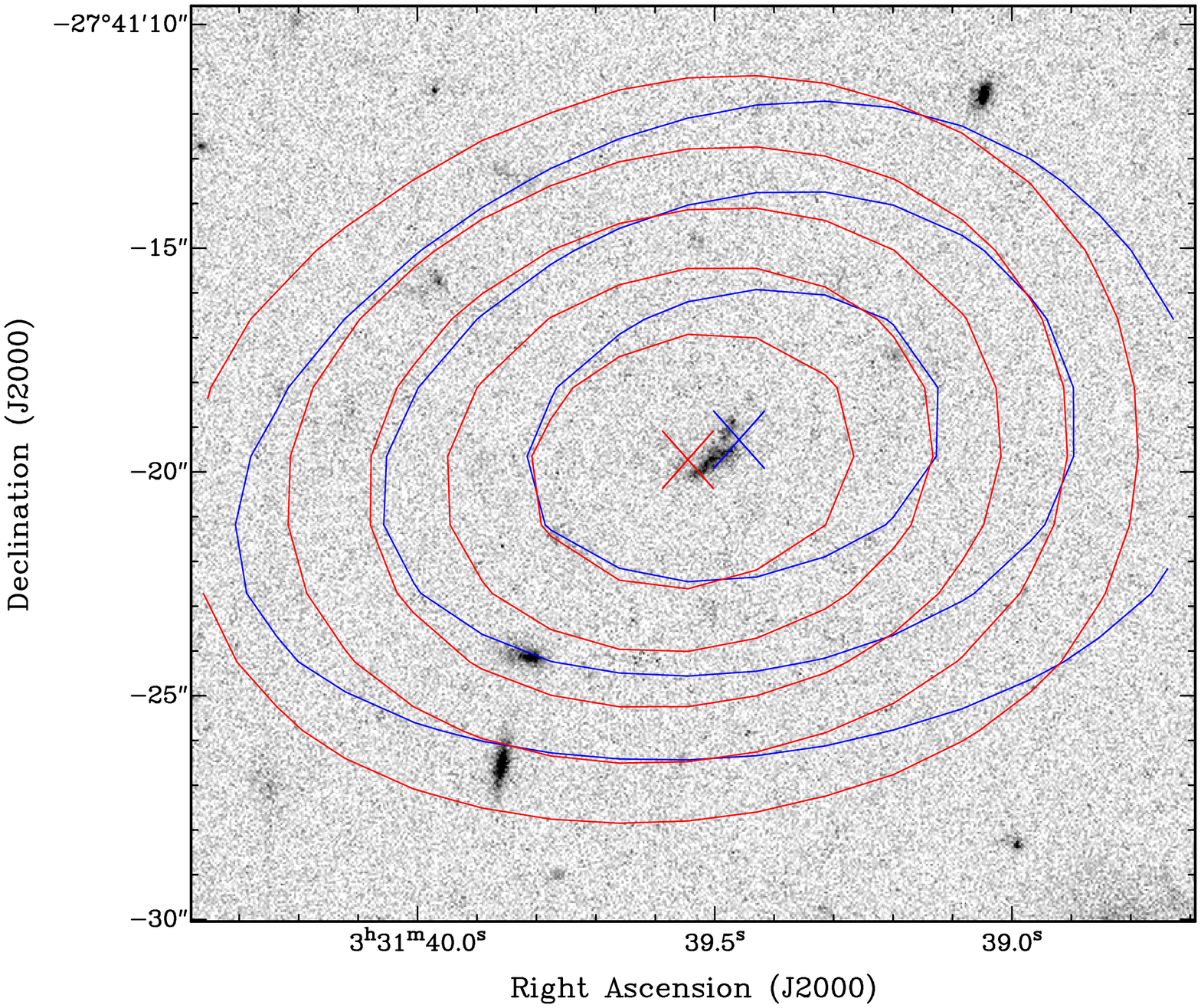}
\hspace{15mm}
\includegraphics[width=6.cm,angle=270]{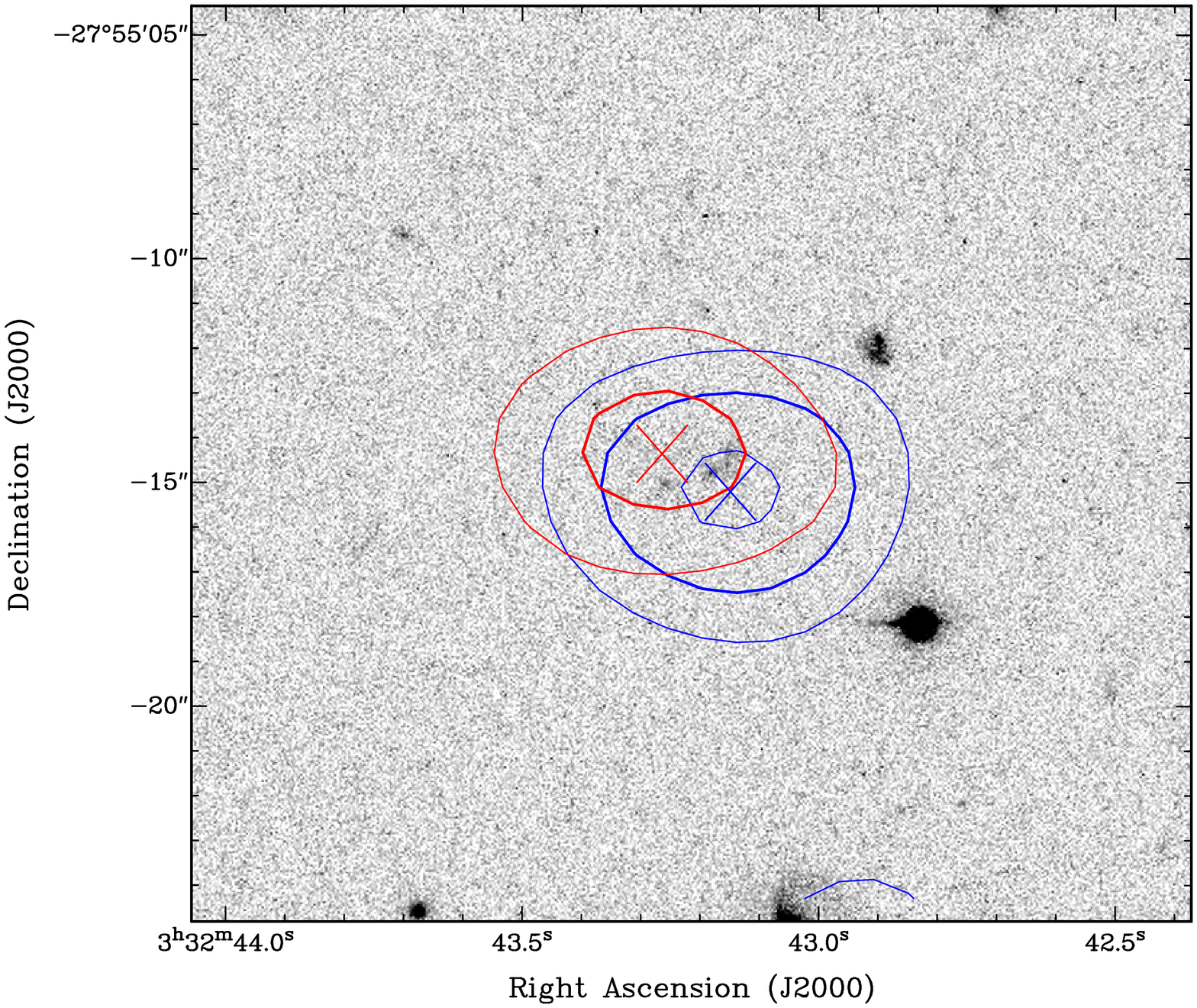}
\caption{The CO(1-0) velocity structure of ALESS 122.1 (left) and ALESS 67.1 (right), overlaid on HST $z$ band (F850LP) images from \protect\cite{rix2004}\protect\footnotemark[3]. 
Both optical counterparts are clumpy or complex in the rest-frame UV, indicating either structured dust or a possible merging system.
Blue contours indicate the CO(1-0) emission integrated from $-300$ to 0 km s$^{-1}$ at 3, 5, 7 .. $\times \; \sigma$. Red contours indicate the CO(1-0) emission integrated from $0$ to +300 km s$^{-1}$ at 3, 5, 7 .. $\times \; \sigma$. Red and blue crosses mark the centroid of the blueshifted and redshifted emission, respectively. The images are 20 $\times$ 20 arcsec. }
\label{fig:redblue}
\end{figure*}

\subsection{Star Formation Efficiency}

The detection of CO(1-0) provides an opportunity to examine the efficiency with which the molecular gas is being converted into stars in these systems. 
Star formation efficiency is commonly defined as SFR/$M$(H$_2$), the inverse of the gas depletion time scale. 
A useful approach is to investigate the star formation efficiency by comparing the two observable quantities, total infrared luminosity $L_{\rm FIR}$ and line luminosity $L'_{\rm CO(1-0)}$. Using these observables instead negates any offsets in SFR/$M$(H$_2$) that may be introduced by the application of an inappropriate CO-to-H$_2$ conversion factor $\alpha_{\rm CO}$. The slope of the $L_{\rm FIR}$ --  $L'_{\rm CO(1-0)}$ relation describes the relationship between the infrared luminosity due to star formation and the total gas content of the system. It is thus a basic observational form of the Kennicutt-Schmidt (K-S) relation between the gas reservoir and the SFR of a system. 

The original form of the K-S relation was established using {\sc H{i}} and CO(1-0) measurements of galaxies in the local universe \citep{schmidt59,kennicutt89,kennicutt98b} and the SFR density was found to be related to gas surface density by a power law, $\Sigma_{\rm SFR} \propto \Sigma_{\rm gas}^N$, with the slope $N$ = 1.4 $\pm$ 0.15 \citep{kennicutt98b}. Using the COLD GASS sample of about 350 local massive galaxies, \cite{saintonge2012} find a global K-S relation of $N$ = 1.18 $\pm$ 0.24. At higher redshifts ($z \sim 1$ -- 3) \cite{genzel2010} suggested that  ``normal'' star forming galaxies show a K-S relation of $N$ = 1.17 $\pm$ 0.09 and SMGs show a similar slope of $N$ = 1.1 $\pm$ 0.2 but offset by 1 dex above the ``normal'' star forming galaxies. Their result comes from high-$J$ (mostly $J$ = 3) transitions of CO, however, and different $\alpha_{\rm CO}$ are adopted for the different populations included in their study. 

The `integrated' K-S relation, as opposed to the original `surface density' K-S relation, is also observed, but assumes the star formation (e.g. as measured by the far-infrared), has the same spatial extent as the molecular gas. 
In the $L_{\rm FIR}$ and  $L'_{\rm CO(1-0)}$ plane ``normal'' star forming galaxies show a slope\footnote[2]{Here slope is $a$, where $L_{\rm FIR}$ $\propto$  $L'_{\rm CO(1-0)}$$^{ a}$} of 1.15 $\pm$ 0.12 and SMGs have a similar slope but offset on average by a factor of four above the relation for ``normal'' star forming galaxies \citep{genzel2010}. Since the $L'_{\rm CO(1-0)}$ luminosities of the ``normal'' star forming galaxies extended into the bright SMG regime ($\gtrsim  10^{10}$ K km s$^{-1}$ pc$^2$) \cite{genzel2010} argued this is evidence that the merging population (ULIRGs and SMGs) are offset from the ``normal'' $z = 0$ disc galaxy population, rather than a change of slope occurring at high luminosities.  More recent work on (local) ULIRGs and SMGs found a $L_{\rm FIR}$ -  $L'_{\rm CO(1-0)}$ relation slope of 1.20 $\pm$ 0.09 for the populations combined, and 1.27 $\pm$ 0.08 and 1.08 $\pm$ 0.14 for the ULIRG and SMG populations, respectively \citep{bothwell2013}. These studies relied on $J$ = 2 or higher transitions of CO converted to CO(1-0), introducing systematic uncertainties in the relation. \cite{ivison2011} used mostly CO(1-0) transitions and fully self-consistent $L_{\rm FIR}$ measurements to find a shallower slope of $0.65 \pm 0.13$ for SMGs, suggesting high transitions of CO may artificially steepen the slope.  

With new CO(1-0) detections of SMGs and BzKs over the last few years, and now our work, we can make a relatively unbiased comparison to local samples. In Figure \ref{fig:sfe} we show  $L_{\rm FIR}$ and  $L'_{\rm CO(1-0)}$, populated with local ULIRGs from \cite{solomon1997} and \cite{chung2009} and local LIRGs from \cite{papadopoulos2012}. For ``normal'' star forming galaxies at $z \sim 2$ we include eight BzK galaxies with published CO detections \citep{daddi2010,aravena2010,aravena2012}, all but three in CO(1-0). Three BzKs have only CO(2-1) detections, which \cite{daddi2010} correct to $L'_{\rm CO(1-0)}$ adopting $r_{21} = 0.84$. This is also the median $r_{21}$ ratio found by \cite{bothwell2013}, so we include the \cite{daddi2010} CO(2-1) detections with the corrected $L'_{\rm CO(1-0)}$. SMGs with CO(1-0) detections come from 
\cite{greve2003}, \cite{ivison2011}, \cite{swinbank2011}, and \cite{harris2012}. Because of the small number of SMGs detected in CO(1-0) (21 including this work) we add seven \cite{bothwell2013} SMGs with CO(2-1) detections, where $L'_{\rm CO(1-0)}$ is derived using $r_{21} = 0.84$. Care was taken in the compilation to use a consistent definition of $L_{\rm FIR}$ across the samples. In the literature, total infrared luminosity $L_{\rm FIR}$ is usually the luminosity across 8 -- 1000 $\mu$m, but in some cases, especially {\it IRAS} samples, authors have given the luminosity across 42.5 -- 122.5 $\mu$m. In these cases (\citealp{solomon1997,chung2009}) we convert the quoted infrared luminosity to $L$(8 -- 1000 $\mu$m) using a factor of 1.9, consistent with typical infrared SEDs (e.g. \citealp{helou1988,ce2001}). 

A linear relation of the form $\log L_{\rm FIR}= a \log L'_{\rm CO(1-0)} + b$ was fit with chi-square minimisation, using the MPFITEXY routine \citep{williams2010} to take into account errors in both coordinates. 
The slopes, $a$, were determined for the various samples, and we report these in the legend of Figure \ref{fig:sfe}. The local ULIRGs and SMGs have consistent slopes of $a = 0.67\pm 0.13$ and $a = 0.61 \pm 0.13$, respectively. The more normal star forming galaxy samples, local LIRGs and BzKs, show marginally steeper slopes of $a = 0.84 \pm 0.09$ and $a = 0.87 \pm 0.61$, but the BzK slope is highly uncertain due to the small number of sources in that sample and the limited luminosity range spanned. The combined BzK and LIRG sample exhibits a slope of $a = 0.88 \pm 0.08$, which is steeper than the ULIRG and SMG combined slope of $a = 0.52 \pm 0.05$ at about the 2$\sigma$ significance level. While a larger sample of SMGs and BzKs with well determined line luminosities $L'_{\rm CO(1-0)}$ is required to definitively say that ULIRGs and SMGs populate a different sequence in the K-S relation to the more normal LIRGs and BzKs, this work does hint that there may be a difference in their SF modes. We stress that our analysis is free of the biases from high $J$ CO transitions which has been present in earlier work (e.g. \citealp{greve2005,daddi2010,genzel2010}). By using the CO(1-0) transition (and CO(2-1) for a small number of sources) we remove the uncertainties in the line luminosity introduced by excitation conditions of the gas. 

\begin{figure*}
\includegraphics[width=10.5cm]{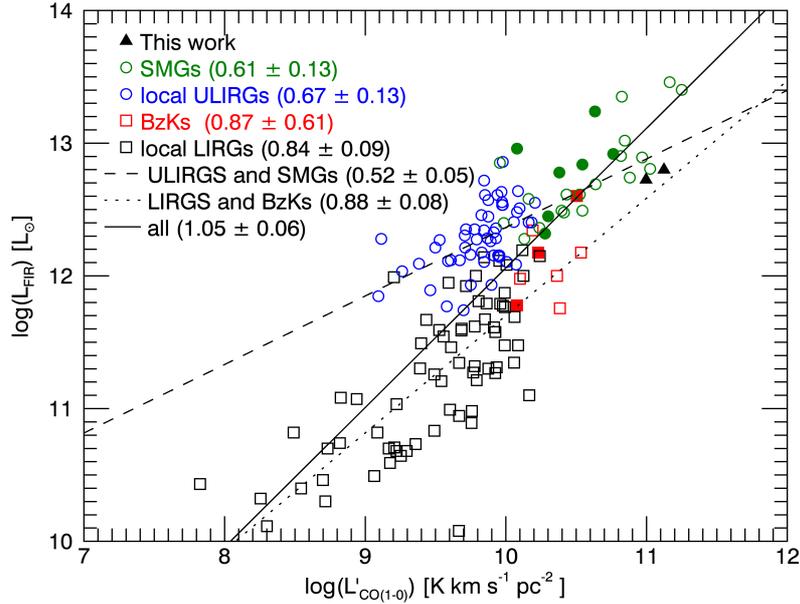}
\caption{$ L_{\rm FIR}$ vs $L'_{\rm CO(1-0)}$ for SMGs, local (U)LIRGs and BzK galaxies. All sources but three BzKs (filled squares) and seven SMGs (filled circles) have robust CO(1-0)  measurements. The filled squares and circles mark sources with CO(2-1) detections which are converted to CO(1-0) using $r_{21} = 0.84$. Linear relations are fitted to the local ULIRGs plus SMGs (dashed) and local LIRGs plus BzK galaxies (dotted), as well as all the samples combined (solid). The numbers in brackets indicate the best fit slope for the different populations.}
\label{fig:sfe}
\end{figure*}

The total infrared luminosities of ALESS 122.1 and ALESS 67.1 \citep{swinbank2014} correspond to star formation rates (SFR) of 940$^{+60}_{-80}$ and 790$^{+100}_{-190}$ M$_\odot$ yr$^{-1}$, respectively, using the infrared conversion from \cite{kennicutt2012}. The gas depletion timescales, $M$(H$_2$)/SFR, are 140 $\pm$ 30 and 130 $\pm$ 40 Myr, which are similar to other SMGs at $z \sim 2$ (e.g. \citealp{bothwell2013}). Assuming little further gas infall from the surrounding environment and 100\% efficiency in converting the gas to stars, the star formation is effectively shut off at $z \sim 2$ and this galaxy would appear as a `red and dead' elliptical by $z \sim 1.5$ (1 Gyr after gas depletion). The estimated total baryonic masses ($>1 \times$ 10$^{11}$ M$_\odot$) and timescales involved are consistent with these SMGs being progenitors of today's massive ellipticals, which has been the consensus in the literature for some time (e.g. \citealp{lilly1999, simpson2014}). 

\footnotetext[3]{https://archive.stsci.edu/prepds/gems/}

\subsection{Dust to Gas Mass Ratios}

As early as the 1980s it was suggested that the emission from dust could be used to measure the mass of the ISM in a galaxy \citep{hildebrand1983}. Dust mass estimates can be determined from the infrared-submm luminosity on the Rayleigh-Jeans tail of the dust SED, and then an appropriate gas-to-dust ratio can be applied to derive the total mass of the molecular ISM \citep{eales2012,scoville2014}. {\em Herschel} surveys have catalogued hundreds of thousands of galaxies (e.g. Herschel-ATLAS, \citealp{eales2010, valiante2016}) in the far-infrared, and SCUBA2 has detected thousands of galaxies in the submm (SCUBA2 Cosmology Legacy Survey, \citealp{geach2016}). Individual CO measurements cannot feasibly be made for all of these galaxies, but dust mass estimates provide a potential avenue for crudely estimating total gas masses for very large samples of galaxies. 

The Milky Way has a gas-to-dust ratio of $\delta_{\rm GDR}$ $\sim$ 130 \citep{jenkins2004}, while the {\it Spitzer} Infrared Nearby Galaxy Survey (SINGS) of 13 local star forming galaxies have $\delta_{\rm GDR} = 130 \pm 20$ \citep{draine2007}. Combining the CO-derived gas measurements from \cite{bothwell2013} and dust mass estimates from far-infrared SEDs by \cite{magnelli2012}, the average $\delta_{\rm GDR}$ for SMGs is estimated to be 90 $\pm$ 25 \citep{swinbank2014}. Combining the dust mass estimates of ALESS 122.1 and ALESS 67.1 from \cite{swinbank2014} with our CO(1-0) gas mass estimates results in gas-to-dust ratios $\delta_{\rm GDR}$ = 170 $\pm$ 30 and 140 $\pm$ 30, respectively. This suggests ALESS 122.1 has an unusually large gas-to-dust ratio, probably due to its extremely large gas reservoir. 

The monochromatic submm flux density has been explored as a probe of gas mass. It potentially provides a simple and effective estimate of both the dust and gas mass content of a galaxy. {\sl Planck} measurements of the submillimeter emission from Milky Way regions yields a Galactic constant of proportionality between 850 $\mu$m luminosity and ISM mass of $\alpha = L_{850}/M_{\rm ISM} = 0.79 \times 10^{20}$ erg s$^{-1}$ Hz$^{-1}$ M$_\odot^{-1}$, while an empirical calibration using 12 local ULIRGs and SINGS survey galaxies yields $\alpha = 1.0 \pm 0.2  \times 10^{20}$ erg s$^{-1}$ Hz$^{-1}$ M$_\odot^{-1}$ \citep{scoville2014}. We convert the measured ALMA 870$\mu$m continuum flux densities of ALESS 122.1 and ALESS 67.1 \citep{hodge2013} to 870 $\mu$m restframe luminosities following \[ L_{870} = 4 \pi D^2 (1 + z) S_{870} K_{\rm corr}, \] where $L_{870}$ is the luminosity in W Hz$^{-1}$, $D$ is the comoving distance, $S_{870}$ is the observed flux density at 870 $\mu$m, and $K_{\rm corr}$ is the K-correction which is given by 
\[ K_{\rm corr} = \left( \frac{\nu_{{\rm obs}}}{\nu_{{\rm obs}(1+z)}} \right) ^{3 + \beta} \frac{e^{h \nu_{{\rm obs}(1+z)}/ k {\rm T}} - 1}{e^{h \nu_{{\rm obs}}/ k {\rm T}} - 1}, \] 
where $\nu_{\rm obs}$ is the observed ALMA frequency (350 GHz),  $\nu_{\rm obs(1+z)}$ is the rest-frame frequency, ${\rm T}$ is the dust SED effective temperature, and $\beta$ is the dust emissivity index. We use temperatures of 32K and 31 K for ALESS 122.1 and ALESS 67.1, respectively, which have been determined from the detailed far-infrared SED fitting of \cite{swinbank2014}. The emissivity index $\beta$  can range from 1.5 to 2.0, but for consistency with \cite{scoville2014} we take $\beta$ = 1.8. We thus find $L_{850}$ = (1.0 $\pm$ 0.1) $\times$ 10$^{31}$ erg s$^{-1}$ Hz$^{-1}$ for ALESS 122.1 and $L_{850}$ = (1.3 $\pm$ 0.1) $\times$ 10$^{31}$ erg s$^{-1}$ Hz$^{-1}$ for ALESS 67.1. This corresponds to monochromatic 850$\mu$m luminosity to mass ratios of (0.75 $\pm$ 0.11) and (1.3 $\pm$ 0.3) $\times$ 10$^{20}$  erg s$^{-1}$ Hz$^{-1}$ M$_\odot^{-1}$ for ALESS 122.1 and ALESS 67.1, respectively, using our measured CO molecular gas masses. This is consistent with the result from \cite{scoville2014} given that the uncertainty in the luminosity-to-mass ratios includes only the flux uncertainties for the CO and ALMA measurements. The luminosity-to-mass ratio given by \cite{scoville2014} includes {\sc H{i}} masses, which we do not include here, and that will also add some uncertainty to the ratio.

\subsection{Radio Thermal Free-Free Emission and the Star Formation Rate}

Most radio measures of star formation use non-thermal radio synchrotron radiation (e.g. \citealp{haarsma2000,seymour2008}), which is emitted by cosmic ray electrons propagating through a galaxy's magnetic field after being initially accelerated by core-collapse supernovae. This synchrotron emission is a complex tracer of star formation and is potentially affected by the inverse-Compton losses against the CMB, which scales as $(1 + z)^4$. The thermal free-free radio emission from ionized H{\sc II} regions traces the massive young stars ($>$5M$_\odot$) which are capable of photoionizing the ISM. As such, it is a more direct tracer of the star formation rate of a galaxy than synchrotron radio emission, and has the same advantage of being a dust-unbiased indicator. The thermal fraction at GHz frequencies for a typical star forming galaxy is $\lesssim$10\% but thermal emission dominates by about 30 GHz  \citep{condon1992}. 

ALESS 122.1 is detected at 1.4 GHz \citep{miller2013} and 5.5 GHz \citep{huynh2015}, with flux densities of 202.6 $\pm$ 14.1 and 71 $\pm$ 9 $\mu$Jy, respectively. Combined with our high frequency 38.1 GHz continuum detection, we can make an estimate of the the amount of radio free-free emission  in ALESS 122.1. The free-free luminosity is derived by fitting the radio SED with a fixed contribution from dust,  thermal free-free radio emission with a fixed slope of $\alpha = -0.1$ ($S \propto \nu^\alpha$), and a non-thermal synchrotron radio component with a fixed slope of $\alpha = -0.8$. The dust contribution is determined from a grey body SED ($S \propto B(\nu,T) \nu^{\beta}$), where $B(\nu,T)$ is the Planck blackbody and $\beta$ is the emissivity index. We set the blackbody function to a temperature of 32K, as determined by \cite{swinbank2014} for ALESS 122.1, and adopt $\beta = 1.5$. The free-free and synchrotron radio components are then allowed to scale to fit the 3 radio continuum data points. The resulting best-fit decomposition of the ALESS 122.1 SED, shown in Figure \ref{fig:aless122_sed}, has a reduced $\chi^2$ value of $\sim$2, indicating a reasonable fit.

The contributions to $S_{\rm 38.1 GHz}$ from dust, free-free and synchrotron radio emission are presented in Table 2.  Using \cite{condon1992}, we determine SFR ($M > 5{\rm M}_\odot$) = 440 $\pm$ 220 $\rm {M}_\odot$ yr$^{-1}$ from the fitted radio free-free emission. 
Using an IMF similar to \cite{kennicutt2012}, we account for stars with $1 < M < 5 {\rm M}_\odot$ using a Salpeter IMF and low mass stars ($0.1 < M < 1{\rm M}_\odot$) with a shallower Kroupa IMF, to find a radio free-free total SFR of 1400 $\pm$  700 ${\rm M}_\odot$ yr$^{-1}$. This is consistent with SFR of 940$^{+60}_{-80}$ ${\rm M}_\odot$ yr$^{-1}$ derived from the infrared luminosity, showing that radio free-free emission has the potential to be a powerful measure of galaxy SFRs.  The fitted thermal fraction of the radio emission from ALESS 122.1 is 73 $\pm$ 37\%, 47 $\pm$ 23\% and 26 $\pm$ 13\% at 38.1, 5.5 and 1.4 GHz, respectively, i.e. 115, 16.6 and 4.2 GHz restframe. This is consistent with the thermal fraction found in M82 \citep{condon1992} and some $z < 0.5$ ULIRGs \citep{galvin2016}. 
This result is from fitting to only three radio data points, and has a large uncertainty, but it shows the potential of radio free-free emission in measuring the star formation rates of galaxies at these redshifts. A caveat however is that an X-Ray stacking analysis by \cite{wang2013} found that ALESS 122.1 is likely to contain an AGN. An additional flat spectrum AGN component can not be ruled out with the current radio data.

Note that the same analysis cannot be performed on ALESS 67.1 because it does not have a 5.5 GHz detection. 

\begin{figure}
\includegraphics[width=8.1cm]{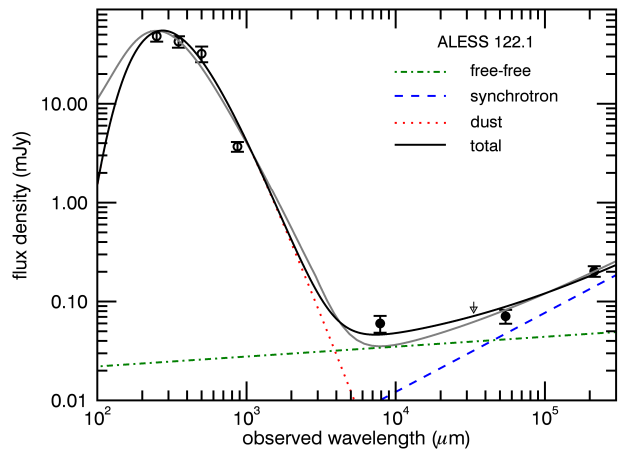}
\caption{The far-infrared to radio SED of ALESS 122.1. We plot {\it Herschel} and ALMA far-infrared and submm datapoints (open circles), and the 1.4 GHz (Miller et al. 2013), 5.5 GHz (Huynh et al. 2015) and 38.1 GHz (this work) radio data (fill circle). 
The downward arrow denotes the 9.0 GHz 4$\sigma$ limit (Huynh et al. in prep). The thermal dust emission component with a temperature of 32K is shown as a red dotted line. The radio free-free (green dot-dashed line) and synchrotron (blue dashed line) components are scaled to produce a best fit to the radio data (solid circles). The black solid line marks the total of all 3 components. For comparison, the best fit SED of Swinbank et al. 2014 is also shown (solid grey line).}
\label{fig:aless122_sed}
\end{figure}

\begin{table}
\centering
\caption{Radio SED fitting results for ALESS 122.1}
\begin{tabular}{llll}  \hline
$S_{\rm 38.1 GHz, total}$ &   $S_{\rm 38.1 GHz, ff}$  & $S_{\rm 38.1 GHz, synch}$  & $S_{\rm 38.1 GHz, dust}$  \\ 
($\mu$Jy) & ($\mu$Jy) & ($\mu$Jy) & ($\mu$Jy) \\ \hline
46 $\pm$ 14 & 34 $\pm$ 17 & 10 $\pm$ 3 & 2.7 (fixed) \\ \hline
\end{tabular}
\end{table}

\section{Conclusions}

We have presented the first results from an ATCA $^{12}$CO(1-0) survey of cold molecular gas from ALMA-detected SMGs in the ECDFS. 

In this first phase we targeted two SMGS at $z \sim 2$. The main results from this work are:
\begin{enumerate}

\item  We detect strong CO(1-0) emission from ALESS 122.1 and ALESS 67.1, which lie at $z = 2.0232$ and $z = 2.1230$, respectively. 
The CO line redshift is consistent with the optical spectroscopic redshift in both cases. The CO emission lines have FWHM $>700$ km s$^{-1}$, which is as expected for more infrared-luminous SMGs. 
The CO(1-0) luminosities are (13 $\pm$ 2) and  (9.9 $\pm$ 1.8) $\times$ $10^{10}$ K km s$^{-1}$ pc$^2$, which correspond to molecular gas masses of 13 and 9.9 $\times$ $10^{10}$ M$_\odot$, for a conversion factor of $\alpha$ = 1.0. 

\item Assuming the CO(1-0) emitting region of ALESS 122.1 and ALESS 67.1 be constrained by the {\sl HST} high resolution optical counterpart, we use the optical source sizes ($<$0.5 arcsec, or 4.2 kpc at $z = 2$) and the measured CO linewidths to estimate dynamical masses of $M_{\rm dyn} \sin^2 i$ $\lesssim$$5 \times 10^{11}$ M$_\odot$ for ALESS 122.1 and ALESS67.1. The spatial offset between the redshifted and blueshifted components of the CO line gives a consistent dynamical mass of $M_{\rm dyn} \sin^2 i$ = (2.1 $\pm$ 1.1) $\times 10^{11}$ M$_\odot$ and (3.2 $\pm$ 0.9) $\times 10^{11}$ M$_\odot$ for ALESS 122.1 and ALESS 67.1, respectively, within 10--15 kpc.  These dynamical masses are similar to the median dynamical masses of typical SMGs \citep{bothwell2013}. 

The stellar masses were combined with the gas mass to derive total baryonic masses of  (1.9 $\pm$ 1.1) $\times$ 10$^{11}$ M$_\odot$ and  (1.4 $\pm$ 0.9) $\times$ 10$^{11}$ M$_\odot$ for ALESS 122.1 and ALESS 67.1, respectively. This implies a gas mass fraction greater than $70$\%.

\item We examine the star formation efficiency of SMGs using the observed  $L_{\rm FIR}$ and  $L'_{\rm CO(1-0)}$ luminosities. We find that ULIRGs and SMGs have a similar slope in $L_{\rm FIR}$ vs $L'_{\rm CO(1-0)}$. Together the ULIRG and SMG population show a slope of 0.60 $\pm$ 0.08, while LIRGs and BzKs show a slightly steeper relation with slope = 0.86 $\pm$ 0.08.  A larger sample with more high redshift CO(1-0) detections is required to definitely conclude that there is a difference in the slope for these populations, but this is some evidence that there is a difference in the SF modes of LIRGS and BzKs versus more extreme ULIRGs and SMGs.

\item We derive gas-to-dust ratios $\delta_{\rm GDR}$ = 170 $\pm$ 30 and  $\delta_{\rm GDR}$ = 140 $\pm$ 30 for ALESS 122.1 and ALESS 67.1, respectively. ALESS 122.1 appears to have an unusually large gas-to-dust ratio, due to its large inferred gas reservoir.  We convert the ALMA submm continuum flux densities to luminosities and find monochromatic 850 $\mu$m luminosity to mass ratios, $L_{850}/M_{\rm gas}$, of (0.75 $\pm$ 0.11) and (1.3 $\pm$ 0.3) $\times$ 10$^{20}$  erg s$^{-1}$ Hz$^{-1}$ M$_\odot^{-1}$ for ALESS 122.1 and ALESS 67.1, respectively, using our measured CO molecular gas masses. 

\item The 38.1 GHz continuum detection of ALESS 122.1 was combined with literature radio data at 1.4 and 5.5 GHz to estimate the free-free radio emission. We find free-free emission makes up  73 $\pm$ 37\% of the radio emission at 38.1 GHz (115 GHz restframe). Converting the free-free emission to SFR yields a SFR of 1400 $\pm$  700 ${\rm M}_\odot$ yr$^{-1}$, consistent with the SFR derived from total infrared luminosity. An AGN contribution to the radio emission can not be ruled out, however. Further sensitive high frequency radio observations between 10 and 100 GHz (restframe) would provide better constraints on free-free radio emission.

\end{enumerate}

CO(1-0) detections are important as the $J$ = 1 transition removes uncertainties in the estimated line luminosities introduced by 
excitation conditions of the molecular gas. Our work has no biases from high $J$ CO transitions, which has hampered earlier work, and illustrates the 
importance of ATCA observations at these frequencies (7mm band, 30 -- 50 GHz). 

The remaining SMGs in the ALESS sample will be targeted with ATCA in the near future as part of our ongoing survey of CO(1-0) from SMGs at $z \sim 2$.
The ALESS sample of SMGs is also being followed-up with ALMA for higher $J$ CO lines and other molecular line tracers of gas such as [C{\sc ii}] and [N{\sc ii}]. The combination of the $J$ = 1  transition CO observations from ATCA with the ALMA observations will reveal the gas excitation conditions and other properties, such as metallicity, in SMGs, shedding even more light on the physical conditions of molecular gas in high redshift galaxies.

\section*{Acknowledgements}
The Australia Telescope Compact Array is part of the Australia Telescope National Facility which is funded by the Australian Government for operation as a National Facility managed by CSIRO.
I.R.S. acknowledges support from the STFC (ST/L00075X/1), the ERC Advanced Investigator Programme DUSTYGAL
(321334) and a Royal Society Wolfson Merit Award.
H.D. acknowledges financial support from the Spanish Ministry of Economy and Competitiveness (MINECO) under the 2014 Ram\'{o}n y Cajal program MINECO RYC-2014-15686.
C.M.C. thanks the College of Natural Sciences at the University of Texas at Austin. 
B.E. acknowledges funding from the European Union 7th Framework Programme (FP7-PEOPLE-2013-IEF) under REA grant 624351.
R.J.I. acknowledges support from ERC Advanced Grant 321302, COSMICISM. 
J.L.W. is supported by a European Union COFUND/Durham Junior Research Fellowship under EU grant agreement number 267209, and acknowledges additional support from STFC (ST/L00075X/1).
W.N.B. acknowledges support from STScI grant HST-GO-12866.01-A. 




\bibliographystyle{mnras}
\bibliography{refs}


\bsp	
\label{lastpage}
\end{document}